\documentclass[aps,prb,reprint,amsmath,amssymb,superscriptaddress,showpacs,floatfix,10pt]{revtex4-1}

\usepackage{graphicx}
\usepackage{dcolumn}
\usepackage{bm}
\usepackage{hyperref}
\usepackage[usenames, dvipsnames]{color}

\begin{document}

\title{Density matrix renormalization group (DMRG) for cyclic and 
centrosymmetric linear chains}

\author{Manoranjan Kumar}
\email{manoranjan.kumar@bose.res.in}
\affiliation{S. N. Bose National Centre for Basic Sciences, Block - JD, Sector - III, Salt Lake, Kolkata - 700098, India}

\author{Dayasindhu Dey}
\affiliation{S. N. Bose National Centre for Basic Sciences, Block - JD, Sector - III, Salt Lake, Kolkata - 700098, India}

\author{Aslam Parvej}
\affiliation{S. N. Bose National Centre for Basic Sciences, Block - JD, Sector - III, Salt Lake, Kolkata - 700098, India}

\author{S. Ramasesha}
\email{ramasesh@sscu.iisc.ernet.in}
\affiliation{Solid State and Structural Chemistry Unit, Indian Institute of Science, Bangalore 560012, India}

\author{Zolt\'an G. Soos}
\email{soos@princeton.edu}
\affiliation{Department of Chemistry, Princeton University, Princeton, New Jersey 08544, USA}

\date{\today}

\begin{abstract}
The density matrix renormalization group (DMRG) method generates the 
low-energy states of linear systems of $N$ sites with a few degrees of 
freedom at each site by starting with a small system and adding sites step by 
step while keeping constant the dimension of the truncated Hilbert space. DMRG 
algorithms are adapted to open chains with 
inversion symmetry at the central site, to cyclic chains and to weakly 
coupled chains. Physical properties rather than energy accuracy is the
motivation. The algorithms are applied to the edge states of linear 
Heisenberg antiferromagnets with spin $S \ge 1$ and to the quantum 
phases of a frustrated spin-1/2 chain with exchange between first and 
second neighbors. The algorithms are found to be accurate for extended Hubbard 
and related 1D models with charge and spin degrees of freedom.
\end{abstract}

\maketitle

\section{Introduction}
Since White~\cite{white92} introduced the density matrix renormalization 
group (DMRG) method and applied it to the spin-1 Heisenberg 
antiferromagnetic chain, the technique has been recognized to be a 
powerful quantitative tool for obtaining the low-energy states of spin 
chains and ladders or of 1D quantum cell models with charge and spin 
degrees of freedom. The reviews of Schollw\"ock~\cite{schollwock2005} and 
Hallberg~\cite{hallberg2006} present the DMRG method in detail. They include 
discussions of infinite and finite DMRG algorithms, of the underlying 
ideas, a careful assessment of approximations, optimization schemes, the 
inclusion of symmetries, and more. The reviews also discuss the scope of 
DMRG applications to diverse 1D systems and the relation of DMRG to other 
numerical and theoretical methods.

\begin{figure}
\begin{center}
\includegraphics[width=\columnwidth]{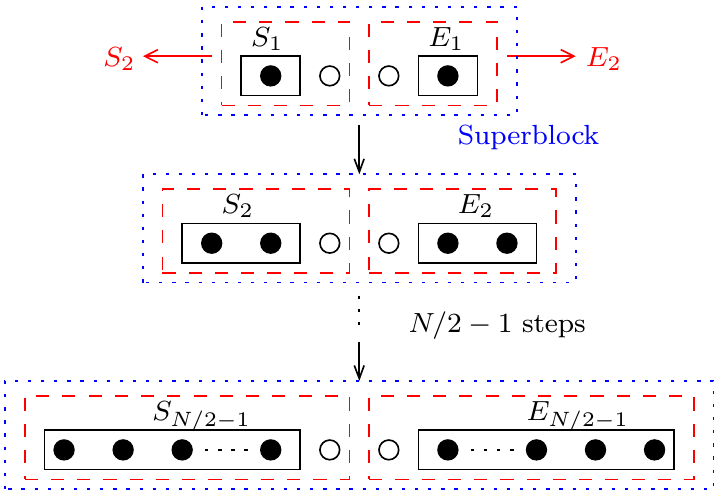}
\end{center}
\caption{Schematic representation of the infinite DMRG algorithm:
Open circles are new sites added at each step to the system (S)
block (left) and environment (E) block (right) until the desired size $N$ is
reached.} \label{fig1}
\end{figure}

The vast majority of DMRG calculations are performed on 1D systems with 
open boundary conditions (OBC) and an even number of sites $N$, as 
proposed by White~\cite{white92} and sketched in Fig.~\ref{fig1}. The 
infinite algorithm increases the targeted superblock size by two sites per 
step. The system (S) and environment (E) blocks are combined into a 
superblock whose Hamiltonian matrix retains the same order independent of 
the superblock's size. Neither the boundary conditions nor even $N$ should 
matter in the thermodynamic limit, $N \to \infty$. In practice, however, 
DMRG calculations are performed on finite systems and the procedure in 
Fig.~\ref{fig1} is not optimal for systems with inversion symmetry at the
central site.

We discuss in this paper other ways for growing 1D systems. The DMRG 
algorithms in Section II retain the key steps of renormalized 
operators, truncation and spanning of the Fock space based on the eigenvalues 
and eigenvectors of the density matrix, and improving the systems block states
by refining the density matrices. We mention three algorithms that differ from 
Fig.~\ref{fig1}. First, open 1D chains with odd $N$ have inversion symmetry 
at the central site and provide complementary information to results for even 
$N$.~\cite{dd1} Second, periodic boundary conditions (PBC) and 
translational symmetry are typically assumed in condensed phases, and DMRG can 
be so modified.~\cite{mk2009} Third, the apparently minor change of adding 
four instead of two sites in Fig.~\ref{fig1} turns out to be important for 
weakly coupled quantum systems in certain topologies.~\cite{mk2010}

DMRG has principally been applied to (a) spin chains or ladders with 
short-range exchange interactions and (b) to extended Hubbard models with 
truncated as well as long-range interactions and to related fermionic 
models with site energies and/or several sites per unit cell. Here we 
discuss spin chains using algorithms that also apply to Hubbard models. 
We consider chains with one spin per unit cell in the thermodynamic 
limit rather than ladders or chains with several spins per unit cell.

The linear Heisenberg antiferromagnet (HAF) is a chain of spin-S sites with 
exchange $J > 0$ between neighbors,
\begin{align}
H_S(N) = J \sum_{r = 1}^{N-1} \vec{S}_r \cdot \vec{S}_{r+1} +
  J_{1N} \vec{S}_1 \cdot \vec{S}_{N}.
\label{eq:hafm}
\end{align}
The open chain has no exchange between sites 1 and $N$ ($J_{1N} =0$), 
while the ring has $J_{1N} = J$. The $S = 1/2$ chain is a 
prototypical many-body problem with known exact properties in the 
thermodynamic limit. Haldane~\cite{haldane83} predicted that integer 
$S$ chains are gapped, as has been confirmed~\cite{white93,schollwock2005,hallberg2006} 
by DMRG and other calculations. HAFs of $S \ge 1$ sites have 
boundary-induced edge states~\cite{sorensen94,tkng94} that are 
discussed in Section III with the conventional DMRG algorithm 
for even $N$ and a recent algorithm for odd $N$.
 
The $J_1-J_2$ model with PBC has spin-1/2 sites, $J_1$ between 
nearest neighbors and antiferromagnetic $J_2 > 0$ between second 
neighbors,
\begin{align}
H(J_1, J_2) = J_1 \sum_r \vec{S}_r \cdot \vec{S}_{r+1} +
 J_2 \sum_r \vec{S}_r \cdot \vec{S}_{r+2}.
\label{eq:j1j2}
\end{align}
The model has been extensively studied: it is frustrated for either 
sign of $J_1$; it has an exact critical point~\cite{hamada88} $J_1/J_2 = -4$ 
between a ferromagnetic and singlet ground state; and a simple exact 
ground state~\cite{ckm69} at $J_1/J_2 = 2$. The quantum phase 
diagram in Section IV has gapped incommensurate spiral phases 
with doubly degenerate singlet ground states and spin correlations of 
finite range as well as gapless phases with nondegenerate ground state 
and quasi-long-range order.~\cite{soos2016} The ground state degeneracy 
in finite PBC systems is between states that are even and odd under 
inversion at sites.  

\section{\label{sec2}Tailored DMRG algorithms}
The general problem is a 1D chain of $N$ sites with $p$ degrees of 
freedom per site. Fermionic systems such as Hubbard models have $p = 4$, 
four states per sites. Spin-$S$ chains have $p = (2S + 1)$ Zeeman levels. 
The dimension of the Fock space is $p^N$. The matrix can typically 
be resolved into sectors with specified symmetries. For example, DMRG 
algorithms usually conserve only $S^z$ in models that conserve the total spin $S$. 
Exact diagonalization (ED) is feasible up to some system size. While it 
is advantageous to work in small sectors, the dimension increases inexorably 
with $N$ and precludes the thermodynamic limit that is often sought. 

As shown in Fig.~\ref{fig1}, two sites are added per step until the desired 
system size $N$ is reached. Let $L$ be the dimension of the Hilbert space
spanned by $H$ in the sector of 
interest and $m$ be the number of states kept in the system block in the 
truncated basis. The DMRG approximation gives constant $L' \ll L$ by 
truncating the Fock space of the system block and renormalizing the operators 
in the system block at each step. The dimension of the DMRG Fock space is $p^2m^2$
and the Hilbert space in a given $S^z$ sector is usually somewhat smaller. When a
superblock of $N$ sites is reached, finite DMRG is performed 
by systematically and repeatedly repartitioning $N$ sites into a larger  ``system" 
block and a smaller ``environment" block, and vice versa. This procedure 
leads to density matrices for different system sizes being obtained from the 
desired eigenstate of the $N$-site superblock. The accuracy, quantified by the 
truncation error,
\begin{align}
P(m) = 1 - \sum_{j=1}^m \omega_j,
\label{eq:trunc}
\end{align}
increases with $m$. The sum is over the eigenvalues $\omega_j$ of 
the density matrix. Typical 
truncation errors are in the range of $10^{-7}$ to $10^{-9}$ for $m \sim 100$ 
to 1000 and can be evaluated for any algorithm.
\begin{figure}[t]
\begin{center}
\includegraphics[width=0.9\columnwidth]{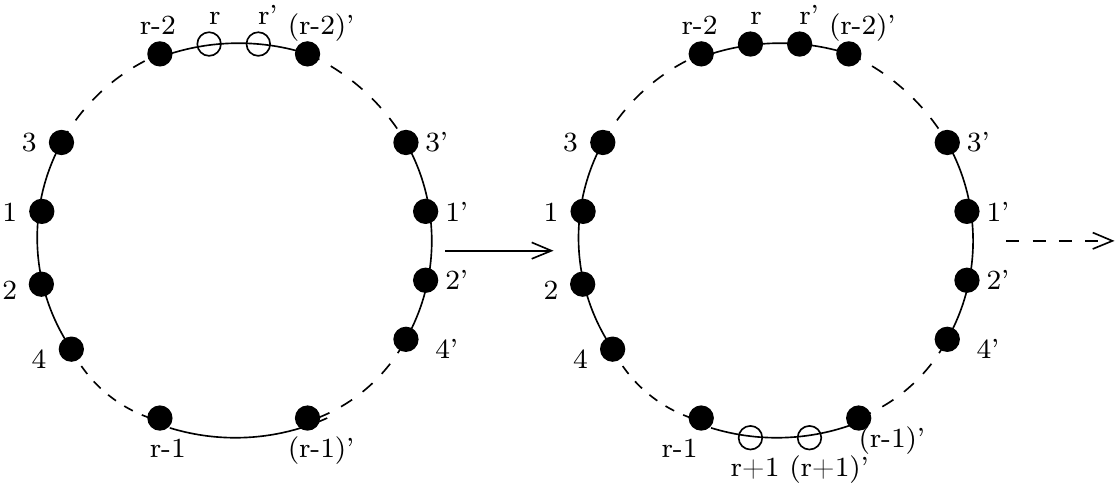}
\end{center}
\caption{Schematic representation of an infinite DMRG algorithm for a PBC 
system: Sites in the left and right blocks are unprimed and primed integers, 
respectively. Filled circles are old sites; open circles are new sites added at 
each DMRG step.} \label{fig2}
\end{figure}

An infinite DMRG procedure~\cite{mk2009} for systems with PBC and even $N$ 
is shown in Fig.~\ref{fig2}.  Two sites are added at each step, alternately in 
the middle of the top and bottom chains. The motivation is symmetry and ground 
state properties. Correlation functions may depend on boundary conditions in 
systems with long-range correlations. Other starting points are possible for 
rings.~\cite{dd2016}
 
\begin{figure}
\begin{center}
\includegraphics[width=\columnwidth]{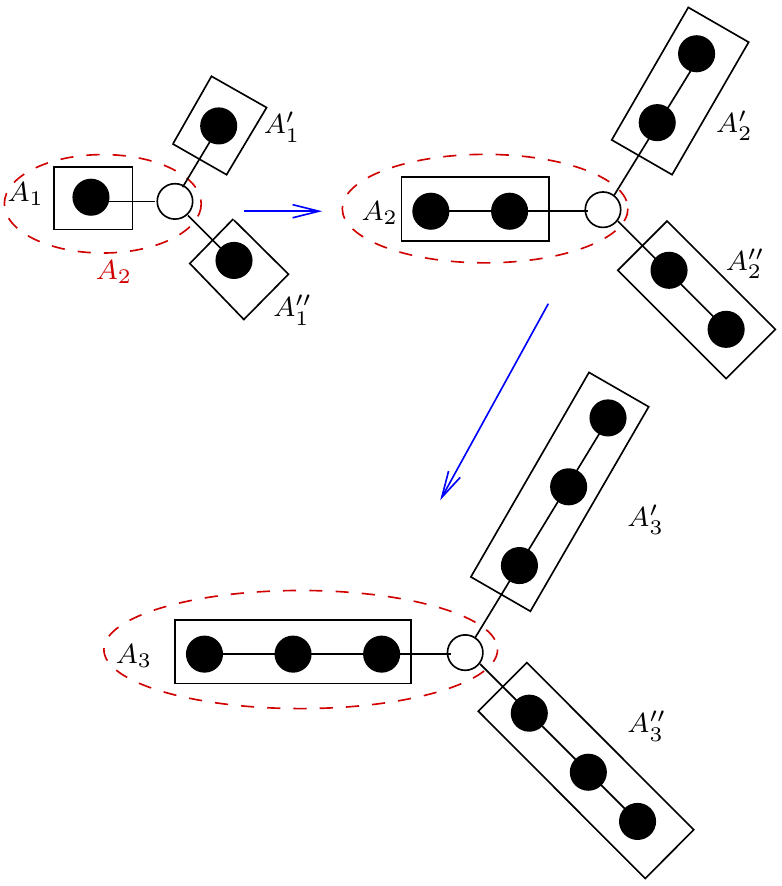}
\end{center}
\caption{Schematic representation of the infinite DMRG algorithm for 
Y junctions with equal arms: At each step, the loop encloses the arm of
the next step and the superblock contains a new site, shown as an open dot, 
and three arms.}
\label{fig3}
\end{figure}

More extensive tailoring of the DMRG algorithm is required for Y junctions,~\cite{mk2016} 
systems of $N = 3n + 1$ sites with three arm of $n$ sites that meet at a central site. 
The infinite algorithm is shown in Fig.~\ref{fig3}. The system is one arm plus the 
central site; the environment is the rest. The junction grows by three sites per step, 
and the system at one step becomes an arm at the next step. The procedure in Fig.~\ref{fig3} 
is immediately applicable to OBC chains with odd $N = 2n + 1$, which can be viewed 
as two arms of $n$ sites and a central site.~\cite{dd1}

Weakly coupled quantum systems are challenging in general and resemble dispersion 
forces in some ways. The $J_1-J_2$ model, Eq.~\ref{eq:j1j2}, discussed in 
Section~IV can be viewed at $J_1 \sim 0$ as two spin-1/2 HAFs on sublattices 
of even and odd numbered sites. The sublattice spin is $S = 0$ when $N/2$ 
is even, $S = 1/2$ when $N/2$ is odd. The conventional DMRG algorithm becomes 
unstable~\cite{white96} for $J_1/J_2 < 1/2$. Adding four instead of two 
spins per step restores the stability for sublattices with $S = 0$ at each 
step.~\cite{mk2010} The change is crucial when $J_1$ is small compared to 
$J_2$ but not when $J_2 < J_1$.

\section{\label{sec3}Edge states of Heisenberg spin chains}
We consider $H_S$ in Eq.~\ref{eq:hafm} and set $J = 1$ as the energy 
unit. The ground state (GS) for OBC has total spin $S_G = 0$ for an 
even number of sites $N$ and $S_G = S$ for an odd number of sites. 
There is no energy penalty for parallel spins at sites 1 and $N$. The 
chain with PBC has $J_{1N} = J$ and $C_N$ translational symmetry. 
Antiferromagnetic coupling leads to the smallest possible GS spin: 
$S_G = 0$ for even $N$ or for integer $S$, and $S_G = 1/2$ for odd 
$N$ and half integer $S$. Exact results in the thermodynamic limit 
for the $S = 1/2$ HAF refer to even $N$. 
 
The energy per site is necessarily the same in the thermodynamic limit, 
but odd $N$ returns $S_G = S$ for OBC and $S_G = 1/2$ for PBC. It 
follows that HAFs with $S \ge 1$ and OBC have edge states that correspond 
to boundary-induced spin density waves (BI-SDWs). The energy gap of edge 
states is~\cite{dd1}
\begin{align}
\Gamma_S(N) = E_0(S, N) - E_0(0, N), 
\label{eq:gap}
\end{align}
where $E_0(S,N)$ is the lowest energy in the sector with total spin $S$. 
Even $N$ leads to $\Gamma_S(N) > 0$. Odd $N$ returns $\Gamma_S(N) < 0$ 
for integer $S$ and $\Gamma_S(N) < 0$ relative to $E_0(1/2,N)$ for half 
integer $S$. Since DMRG algorithms conserve $S^z$ rather than $S$, the 
most accurate results are for the GS in sectors with increasing $S^z$ 
and $\Gamma_S(N) > 0$. The singlet (doublet) for $\Gamma_S(N) < 0$ is an 
excited state in the $S^z = 0$ ($S^z = 1/2$) sector for integer 
(half integer) $S$. 

\begin{figure}
\begin{center}
\includegraphics[width=\columnwidth]{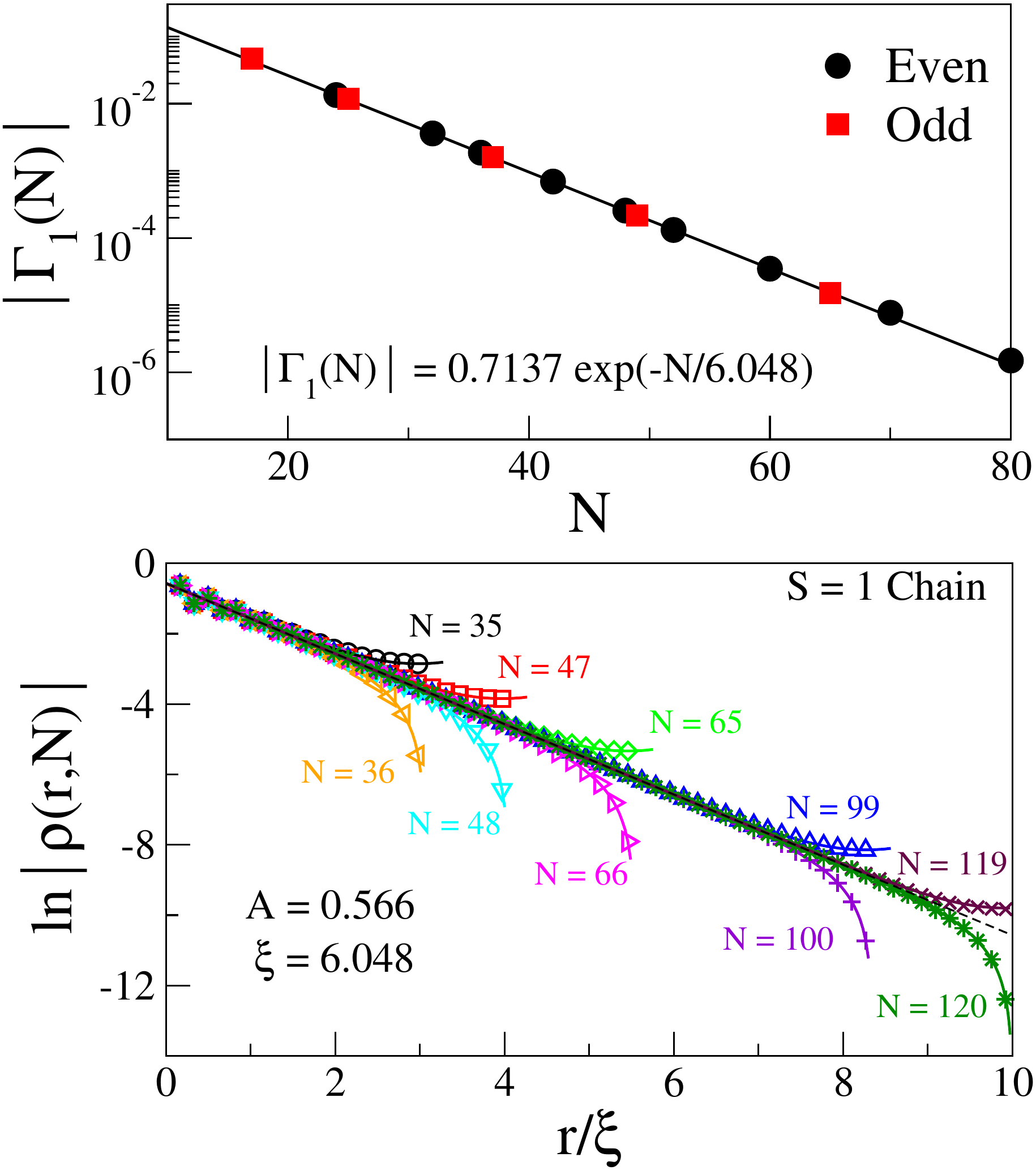}
\end{center}
\caption{Upper panel: Singlet-triplet gap $|\Gamma_1(N)|$ of $S = 1$ 
chains with OBC and $N$ spins in Eq.~\ref{eq:hafm}. Lower panel: DMRG 
results for $|\rho(r,N)|$ to the middle of $S = 1$ chains. Lines are 
Eq.~\ref{eq:bispden} with $\xi= 6.048$ and $A = 0.566$. Even and odd 
$N$ deviate from $A \exp(-r/\xi)$ near the middle of chains.}\label{fig4}
\end{figure}

According to the nonlinear sigma model, the $S = 1$ chain has an 
effective spin $s' = 1/2$ at each end.~\cite{sorensen94} The size 
dependence of the singlet-triplet gap is
\begin{align}
\Gamma_1 (N) = (-1)^N J_e \exp(-N/\xi),
\label{eq:stgap}
\end{align}
where $\xi$ is the spin correlation length in the thermodynamic limit and 
$J_e$ is undetermined. The upper panel of Fig.~\ref{fig4} shows $|\Gamma_1|$
as a function of system size. DMRG returns $\xi= 6.048$ and $J_e = 0.7137$, 
consistent with previous even-$N$ results.~\cite{white93,sorensen94} The 
gap $-\Gamma_1$ for odd $N$ agrees quantitatively with Eq.~\ref{eq:stgap}.  

The spin density at site $r$ is 
\begin{align}
\rho (r,N) = \langle S_r^z \rangle, \qquad r = 1, 2, \ldots, N. 
\label{eq:spden}
\end{align}
The expectation value is with respect to the state of interest in the 
Zeeman sector $S^z = S$. Singlets have $\rho(r,N) = 0$ at all sites. 
SDWs in $S \ge 1$ chains have equal spin density at $r$ and $N+1-r$ by 
symmetry. We model the spin densities of integer S chains as~\cite{dd1}
\begin{align}
\rho (r, N) = A (-1)^{r-1} \left( \exp(-r/\xi) - (-1)^N \right. \notag\\
\left. \times \exp(-(N+1-r)/\xi)\right),
\label{eq:bispden}
\end{align}
where $A$ is an amplitude. The SDWs are in phase for odd $N$, out of phase 
for even $N$. The lower panel of Fig.~\ref{fig4} shows $|\rho(r,N)|$ up to 
the middle of even and odd chains. The lines are Eq.~\ref{eq:bispden} with 
continuous $r$ and $\xi = 6.048$, $A = 0.566$ for all chains.

\begin{figure}
\begin{center}
\includegraphics[width=\columnwidth]{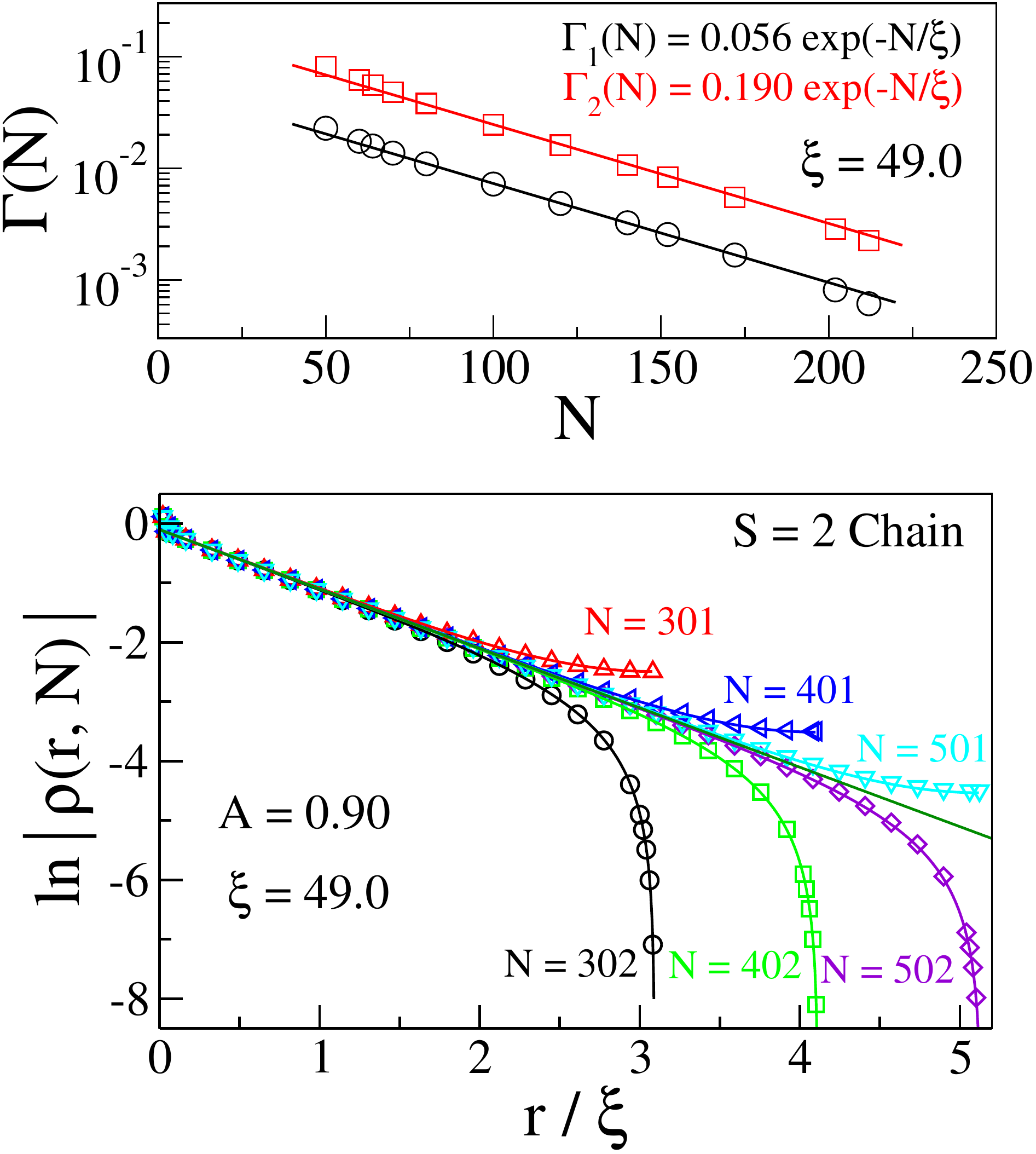}
\end{center}
\caption{Upper panel: Edge-states $\Gamma_1(N)$ and $\Gamma_2(N)$ of 
$S = 2$ chains with OBC and $N$ spins in Eq.~\ref{eq:hafm}. Lower 
panel: DMRG results for $|\rho(r,N)|$ in the $S^z = 2$ sector to the middle 
of $S = 2$ chains. Lines are Eq.~\ref{eq:bispden} with $\xi = 49.0$ and 
$A = 0.90$. The chains deviate from $A \exp(-r/\xi)$ in the middle.} \label{fig5}
\end{figure}

The $S = 2$ chain has smaller Haldane gap~\cite{nakano2009} and hence longer 
correlations. The gaps in Eq.~\ref{eq:gap} are $\Gamma_1$ to the triplet 
($S = 1$) and $\Gamma_2$ to the quintet ($S = 2$). The ratio is 
$\Gamma_2/\Gamma_1 = 3$ for Heisenberg exchange between the effective 
spins $s' = 1$ at the ends. The panels of Fig.~\ref{fig5} show the gaps for 
even $N$ and the quintet spin densities to the middle of even and odd 
chains.~\cite{dd1} The lines have spin correlation length $\xi = 49.0$ and amplitude 
$A = 0.90$ in Eq.~\ref{eq:bispden}. Even and odd $N$ draw attention to the 
relative phases of SDWs, and the analysis of spin densities in the middle 
clarifies the thermodynamic limit. Spin densities are accurately found to 
$N \sim 500$ while the numerical accuracy~\cite{dd1} limits the gaps to 
$N \sim 220$ because Eq.~\ref{eq:gap} involves small differences between 
total energies. The calculated ratio $\Gamma_2/\Gamma_1 = 3.4$ is somewhat 
larger than 3.

\begin{figure}
\begin{center}
\includegraphics[width=\columnwidth]{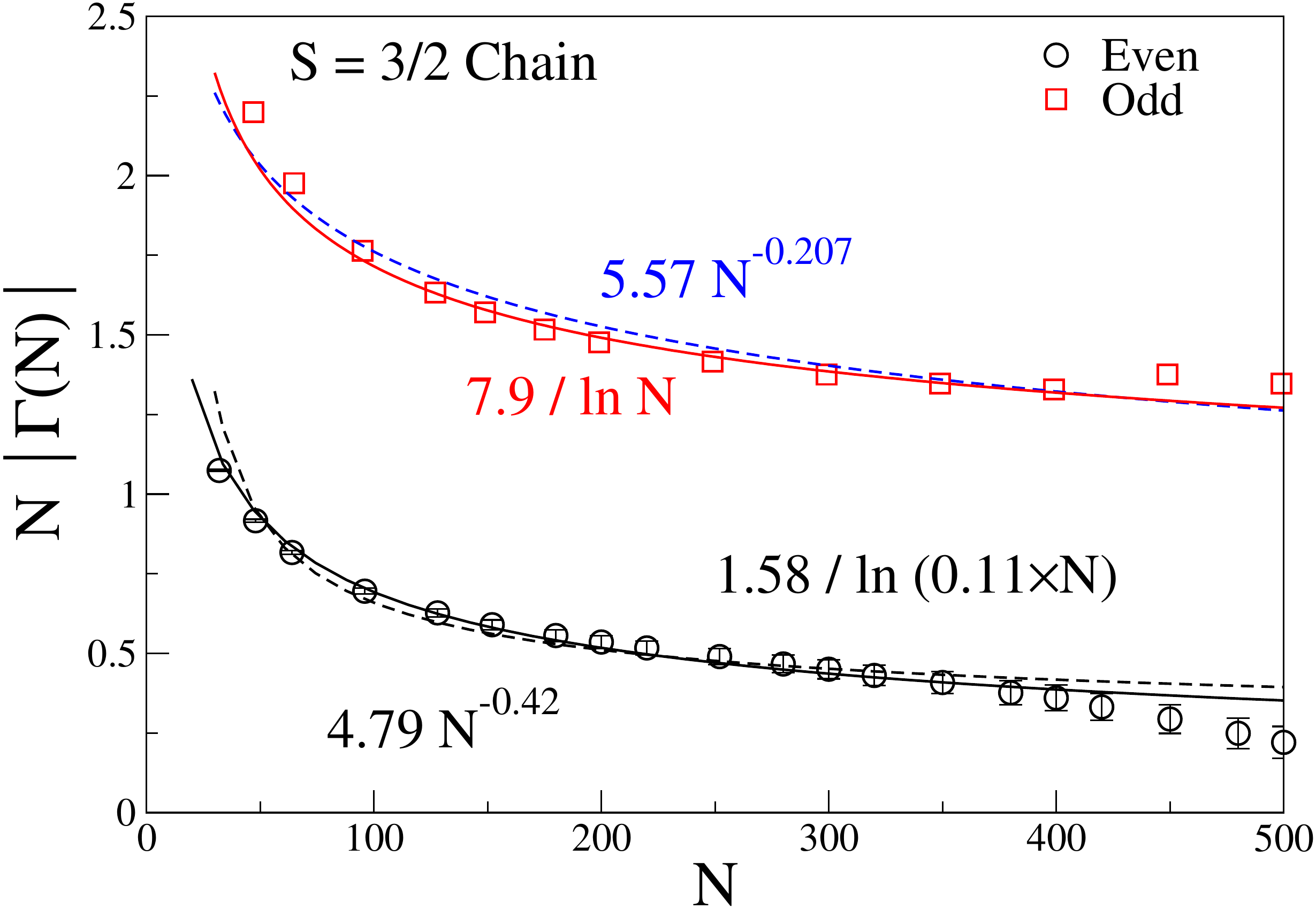}
\end{center}
\caption{DMRG results for the gaps $\Gamma_1(N)$ and $|\Gamma_{3/2}(N)|$ 
for $S = 3/2$ chains with OBC and $N$ spins in Eq.~\ref{eq:hafm}. The 
solid and dashed lines are two parameter fits, with $1.58/\ln 0.11 N$ 
for even $N$ from Ref.~\onlinecite{fath2006}.} \label{fig6} 
\end{figure}

HAF chains with half integer $S \ge 3/2$ are gapless and their edges states 
are fundamentally different. Even chains have a singlet GS while odd chains 
have $S_G = S$ and BI-SDWs with half integer $S > 1/2$. The even $S = 3/2$ 
chain has a gap $\Gamma_1(N)$ that decreases faster than $1/N$ and has been 
studied~\cite{fath2006} to $N = 192$. Odd chains have $-\Gamma_{3/2}(N)$ from 
the quartet GS to the $S = 1/2$ excited state, which is the first excited 
state in the $S^z = 1/2$ sector. Both gaps are shown in Fig.~\ref{fig6} up to 
$N = 450$. The dashed line for even $N$ is the two-parameter fit of 
Ref.~\onlinecite{fath2006}, while the solid line is a two-parameter power law. The 
gap for odd $N$ has larger amplitude and weaker size dependence. 

\begin{figure}
\begin{center}
\includegraphics[width=\columnwidth]{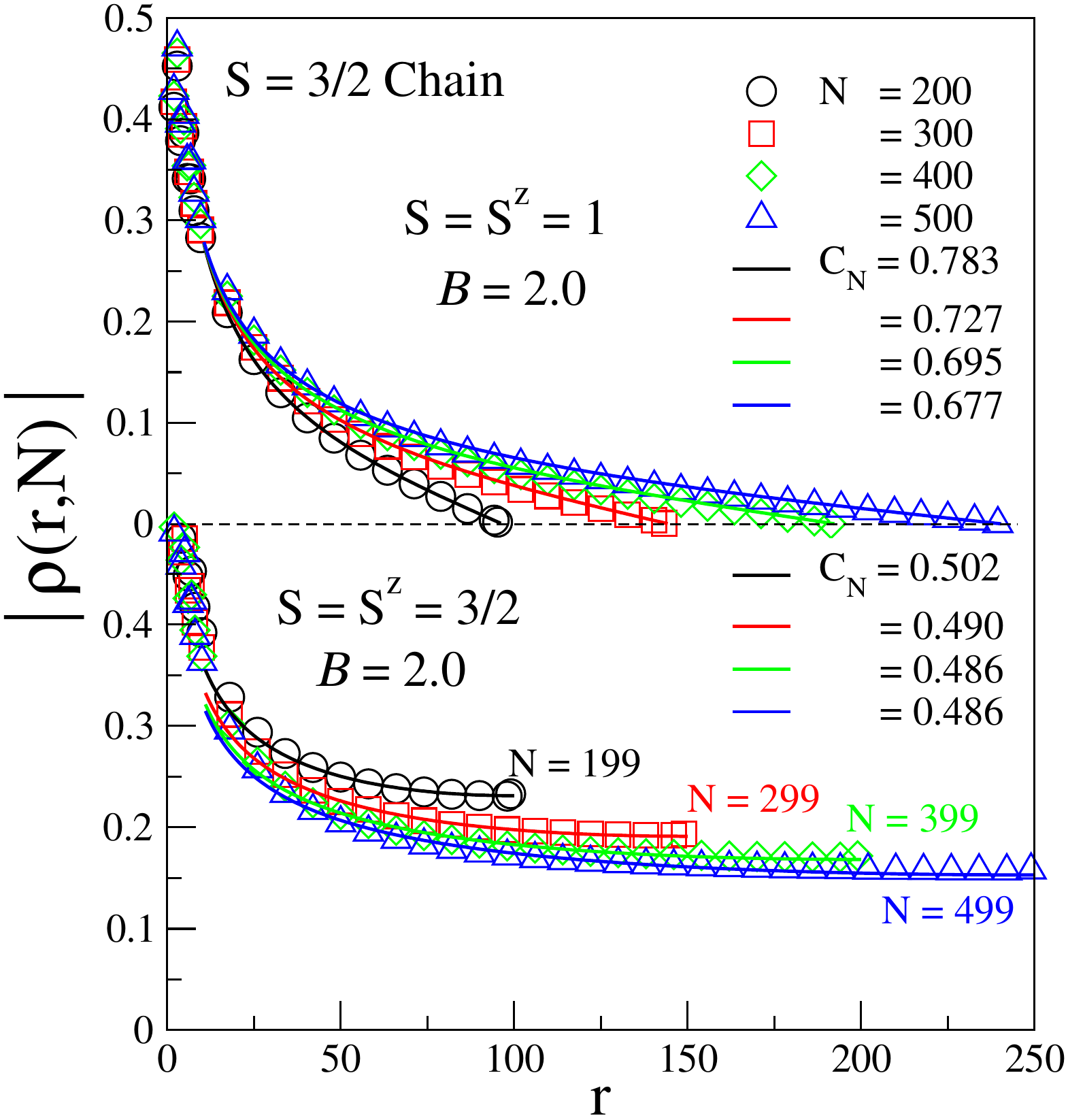}
\end{center}
\caption{DMRG results for $|\rho(r,N)|$ to the middle of $S = 3/2$ chains 
with OBC and $N$ spins in Eq.~\ref{eq:hafm}. The lines are Eq.~\ref{eq:spd_model} 
with $B = 2$ and amplitudes $C_N$.} \label{fig7} 
\end{figure}

The BI-SDWs of the $S = 3/2$ chain are also qualitatively different. 
Figure~\ref{fig7} shows the spin densities to the middle of even and odd 
chains. The total spin density is rigorously 1 for even $N$ and 3/2 for 
odd $N$, as required, but the BI-SDWs are not localized. The sum over all 
sites of the absolute spin density, $|\rho(N)| = \sum_r |\rho(r,N)|$, 
diverges in the thermodynamic limit.

In phase BI-SDWs for odd $N$ result in large amplitude at the middle that 
decreases in Fig.~\ref{fig7} slightly faster than $r^{-1/2}$, while out of 
phase SDWs cancel for even $N$. The spin correlation 
functions~\cite{hallberg96,affleck89} of the $S = 1/2$ and 3/2 HAFs and the 
size dependence of the amplitude suggests~\cite{dd1} modeling the spin 
densities as
\begin{align}
\rho (r, N) = (-1)^{r-1} C_N \left( \left(\frac{\ln B r }{r}\right)^{1/2} - 
(-1)^N \right. \notag \\
\left. \times \left(\frac{\ln B(N+1-r)}{N+1-r}\right)^{1/2} \right).
\label{eq:spd_model}
\end{align}
The amplitude $C_N$ depends on system size because the SDWs are not 
localized. The lines for $|\rho(r,N)|$  in Fig.~\ref{fig7} are 
Eq.~\ref{eq:spd_model} with $B = 2$ and the indicated $C_N$. The spin 
densities are adequately fit in the central region in either case. 
Deviations are limited to $r < 10$ when $N$ is even and to $r < 15$ 
when $N$ is odd, and such deviations are also seen in Figs.~\ref{fig4}b 
and~\ref{fig5}b for the first few sites of integer $S$ chains.

\section{\label{sec4}$J_1-J_2$ model: Incommensurate and decoupled phases}
\begin{figure}
\begin{center}
\includegraphics[width=\columnwidth]{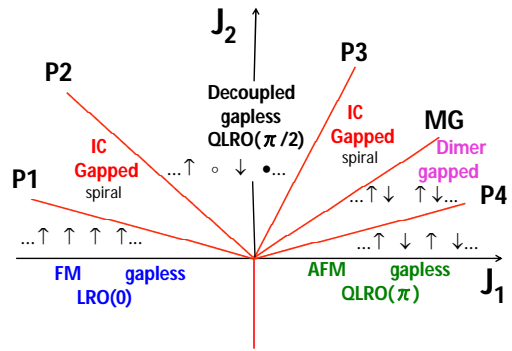}
\end{center}
\caption{Quantum phase diagram of $H(J_1,J_2)$, Eq.~\ref{eq:j1j2}, with 
critical points P1-P4 at $J_1/J_2 = -4$,  $-1.24$, 0.44 and 4.148. 
The exact point P1 is between a gapless ferromagnetic phase and a gapped 
incommensurate (IC) phase. The gapless decoupled phase is between P2 and P3; 
open and closed circles denote spins pointing in and out of the plane. The 
gapped IC phase extends to the MG point, $J_1 = 2J_2$, and the gapped dimer 
phase to P4, beyond which lies a gapless phase with quasi-long-range order.}
\label{fig8}
\end{figure}

The $J_1-J_2$ model, Eq.~\ref{eq:j1j2}, has been studied in several 
contexts. The quantum phase diagram in Fig.~\ref{fig8} has exact critical 
and special points. The thermodynamic limit at $J_2 = 0$, $J_1 > 0$ is a 
spin-1/2 HAF. The gapless phase has a nondegenerate singlet GS and spin 
correlations with quasi-long-range order (QLRO($\pi$)) at wave vector 
$q = \pi$. The ferromagnetic phase with $J_1 < 0$ and LRO(0) extends to 
the exact critical point~\cite{hamada88} $P1 = J_1/J_2 = -4$. The gapless 
phase at $J_1 = 0$ has QLRO($\pi$/2) and corresponds to spin-1/2 HAFs on 
sublattices of even and odd-numbered sites. The exact GS at the 
Majumdar-Ghosh point,~\cite{ckm69} $J_1/J_2 = 2$, are doubly degenerate 
and very simple: They are the two Kekul\'e valence bond diagrams of organic 
chemistry in which adjacent spins are singlet paired, 
$(\alpha \beta - \beta \alpha)/\sqrt 2$.  The gapped dimer phase has 
finite-range correlations at $q = \pi$ and spontaneously broken 
inversion symmetry at sites. The initial studies focused on the critical 
point~\cite{okamoto92, chitra95, eggert96} $P4 = J_1/J_2 = 4.148$ 
where a singlet-triplet gap $E_m$ opens, the GS becomes doubly degenerate 
and range of spin correlations becomes finite.

The wave vector $q_G$ of GS spin correlations evolves from $q_G = 0$ at 
$P1$ to $\pi$ at MG. The gapped incommensurate spiral phases have doubly 
degenerate GS with $\pm q_G$. The discussion so far is not at all 
controversial. There has been disagreement, however, about the critical 
points P2 and P3 that delimit the decoupled phase in Fig.~\ref{fig8}. 
Field theories~\cite{white96,itoi2001,furukawa2012} have restricted the 
phase to the point $J_1 = 0$ using renormalization group flows to 
distinguish between gapped and gapless phases. The singlet-triplet gap 
is very small indeed for $J_1 \sim 0$, far beyond numerical methods, and 
field theory also entails approximations. Another approach to the phases 
at small $J_1$ is to focus on GS degeneracy. The values of P2, P3 in 
Fig.~\ref{fig8} are mainly based on degeneracy.~\cite{soos2016}

Finite $J_1-J_2$ models of $N = 4n$ spins have discrete wave vectors that 
change in steps of $\pi/2n$ between $q = 0$ and $\pm \pi$ in the first 
Brillouin zone. The singlet GS is nondegenerate except at $2n$ values of 
$J_1/J_2$ between $-4$ and 2 where it is doubly degenerate, even and odd, 
$\sigma = \pm 1$, under inversion at sites.~\cite{soos2016} The first and 
last degeneracy are at $J_1/J_2 = -4$ and 2, respectively, for any system 
size. Increasing $J_1/J_2$ generates a staircase of $2n$ steps at which 
$q_G$ changes by $\pi/2n$. 

ED is limited to $N = 28$ in our calculations. To study larger $N = 4n$ 
systems, we evaluate the static structure factor $S(q)$, with $q$ varying 
from 0 to $\pm \pi$ in steps of $\pi/2n$
\begin{align}
S(q) = \sum_r \langle \vec{S}_0 \cdot \vec{S}_r \rangle \exp(-irq).
\label{eq:sq}
\end{align}
The expectation values are in the singlet GS of systems with PBC and 
$-4 \le J_1/J_2 \le 2$. The correlation functions depend only on the 
separation $r$ between spins. The structure factor peaks~\cite{soos2016} 
at $q = q_G$ except in the immediate vicinity of $J_1/J_2 = 2$. 

\begin{figure}
\begin{center}
\includegraphics[width=\columnwidth]{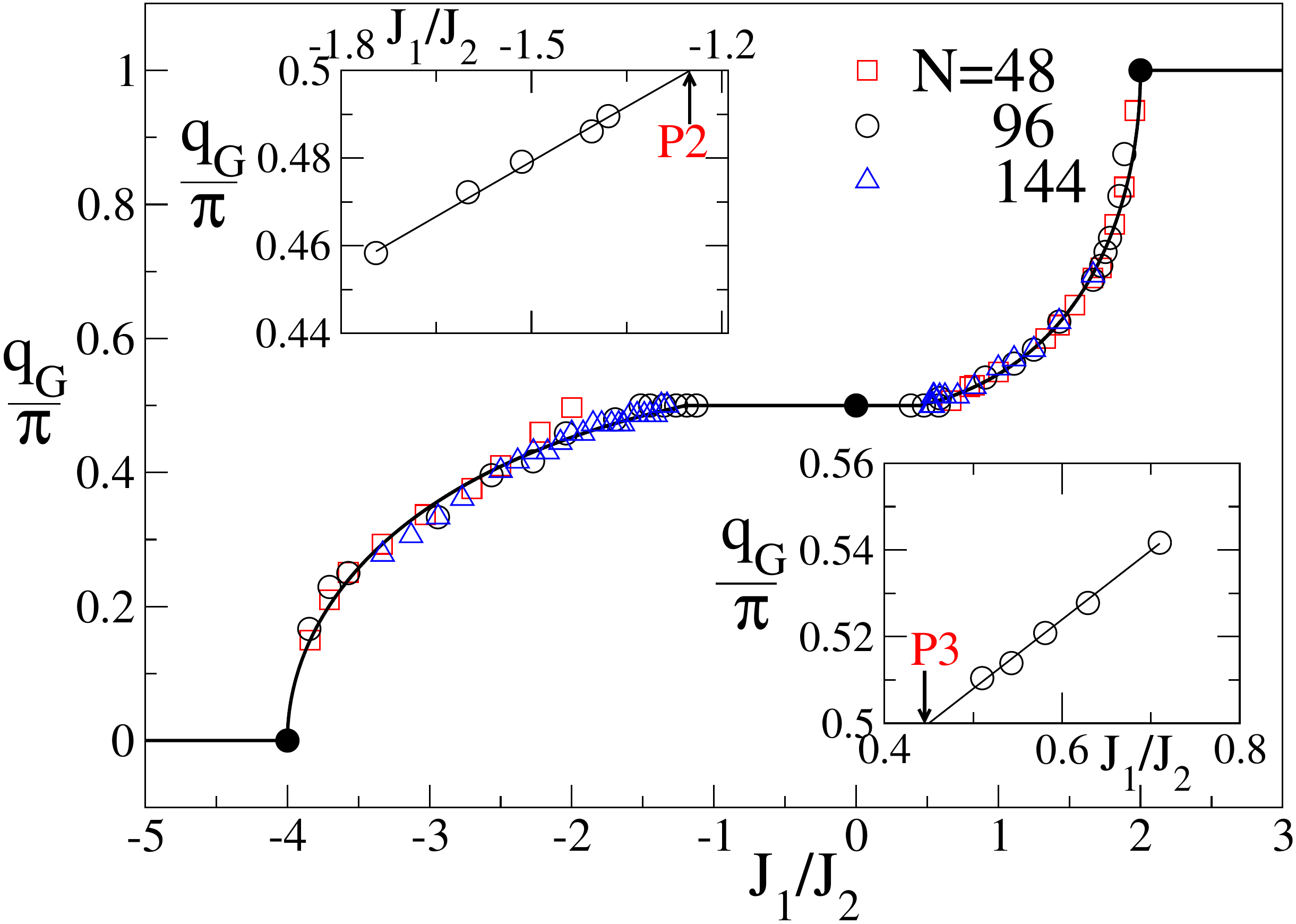}
\end{center}
\caption{DMRG results for the wave vector $q_S$ of GS correlations as 
a function of $J_1/J_2$ in models with PBC and $N = 4n$ spins in 
Eq.~\ref{eq:j1j2}. Closed circles are exact in the thermodynamic limit. 
The insets show to $N = 192$ the $J_1/J_2$ values at which the 
$q_S/\pi$ plateau is reached in a step $2/N$ at $J_1 < 0$ and left at 
$J_1 > 0$; linear extrapolation give the critical points P2 and P3. The 
line is discussed in Ref.~\onlinecite{soos2016}.} \label{fig9}
\end{figure}

DMRG calculations yield $S(q)$ and its maximum $q_G$ as a function of 
$J_1/J_2$. Results to $N = 144$ are shown in Fig.~\ref{fig9}. The line 
is a fit~\cite{soos2016} that takes into account the square-root 
singularities at $-4$ and 2. The $q_G = \pi/2$ plateau is particularly 
important. The insets show up to $N = 192$ when the plateau is reached 
at $J_1/J_2 < 0$ and left at $J_1/J_2 > 0$. Linear extrapolation gives 
the critical points $P2 = -1.24$ and $P3 = 0.44$ in Fig.~\ref{fig8}. 
Finite size and discrete $q$ are advantageous here since $q_G$ is 
exactly $\pi/2$ and the GS is nondegenerate until $q_G$ changes by 
$\pm \pi/2n$. The gapless decoupled phase in the interval 
$-1.24 \le J_1/J_2 \le 0.44$ requires a modest extrapolation when viewed 
in terms of GS degeneracy. 

Okamoto and Nomura~\cite{okamoto92} evaluated P4 by noting that a doubly 
degenerate GS in the dimer phase requires two singlets at lower energy than 
the lowest triplet. ED to $N = 24$ gave $J_1/J_2$ at which the singlet 
excited state and the triplet are degenerate. The weak size dependence of 
the crossing allowed accurate extrapolation to $P4  = 4.148$. ED,~\cite{okamoto92}
DMRG~\cite{chitra95} and field theory~\cite{eggert96} are in excellent agreement 
for $P4$.

ED to $N = 28$ for excited state crossings~\cite{mk2015} also returns 
estimates for $P2$ and $P3$. The size dependence is stronger but entirely 
consistent with critical points based on GS degeneracy. As for $P4$, the 
singlet GS has $q_G = \pi$ for $J_1/J_2 \ge 2$ and $P4$ is related to the 
divergence of $S(\pi)$ in the QLRO($\pi$) phase. Spin correlations are 
limited to nearest neighbors at the MG point, where $S_{\text{MG}}(\pi) = 3/2$ 
is exact. The peak $S(\pi)$ increases for $J_1/J_2 > 2$ as the range of 
spin correlations increases and it diverges at $P4$ in the thermodynamic 
limit.~\cite{mk2015} QLRO($q_G$) phases have divergent $S(q_G)$. 

The spin correlations in Eq.~\ref{eq:sq} were obtained with a DMRG 
algorithm for PBC. The largest separation is $r = 2n$ for $N = 4n$. 
Correlations in OBC systems are typically computed as close to the 
center as possible in order to minimize end effects; $r = 2n$ requires 
sites $n$ and $3n$ in systems of $4n$ spins. Significant end effects 
can be demonstrated in half-filled systems of free electrons. There 
is presumably no problem when $N$ exceeds the range of correlations. 
It is difficult to assess the accuracy of structure factors based on 
OBC spin correlations, and PBC is clearly preferable.

\section{\label{sec5}Conclusions}
As summarized in Section II, there are different ways to grow 1D systems 
with infinite DMRG algorithms. The physics of the system rather than 
energy accuracy is the reason for tailored algorithms. The general DMRG 
methodology holds directly or with minor modification for 
these algorithms. Comparable truncation errors, $P(m)$ in Eq.~\ref{eq:trunc}, 
are expected and found. The scheme in Fig.~\ref{fig1} grows two sites per 
step a 1D chain with an even number of sites. The algorithm developed for 
Y junctions in Fig.~\ref{fig3} also generates a 1D chain, two sites per step, 
with an odd number of sites. The scheme in Fig.~\ref{fig2} grows a cyclic 
1D chain two sites per step. Adding four instead of two sites per steps 
makes the conventional algorithm applicable to weakly coupled spin-1/2 chains. 

\begin{acknowledgements}
 MK thanks the Department of Science and Technology, India for the Ramanujan 
Fellowship SR/S2/RJN-69/2012 for funding computation facility through 
SNB/MK/14-15/137. SR thanks the DST, India for support of this work through various 
grants. ZGS thanks the National Science Foundation, USA.
\end{acknowledgements}

\end{document}